\documentclass[9pt,twocolumn,twoside]{osajnl}
\usepackage{babel}
\journal{ol}
\setboolean{shortarticle}{true} 

\begin{document}

\title{A normal form for frequency combs and localized states in Kerr–Gires–Tournois interferometers}
\author[1,2,*]{Thomas G. Seidel}
\author[2]{Julien Javaloyes}
\author[1,2,3]{Svetlana V. Gurevich}

\affil[1]{Institute for Theoretical Physics, University of M\"unster, Wilhelm-Klemm-Str. 9, 48149 M\"unster, Germany}
\affil[2]{Departament de F\'{\i}sica, Universitat de les Illes Balears \& IAC-3, Cra.\,\,de
	Valldemossa, km 7.5, E-07122 Palma de Mallorca, Spain}
\affil[3]{Center for Nonlinear Science (CeNoS), University of M\"unster, Corrensstraße 2, 48149 M\"unster, Germany}

\affil[*]{Corresponding author: thomas.seidel@uni-muenster.de}
\graphicspath{{./}{./figs/}} 

\begin{abstract}
We elucidate the mechanisms that underlay the formation of temporal localized states and frequency combs in vertical external-cavity Kerr-Gires-Tournois interferometers. We reduce our first principle model based upon delay algebraic equations to a minimal pattern formation scenario. It consists in a real cubic Ginzburg-Landau equation modified by high-order effects such as third order dispersion and nonlinear drift. The latter are responsible for generating localized states via the locking of domain walls connecting the high and low intensity levels of the injected micro-cavity. We interpret the effective parameters of the normal form in relation with the configuration of the optical setup. Comparing the two models, we observe an excellent agreement close to the onset of bistability.
\end{abstract}

\maketitle

Optical frequency combs (OFCs) offer a wide range of application in a variety of fields including high-precision optical spectroscopy~\cite{UHH-NAT-02}, arbitrary waveform generation~\cite{CW-NAP-10} or ultrabroad band coherent optical communications, see~\cite{D-JOSAB-10,PPR-PR-18} for reviews. Prominent realizations of OFCs include mode-locked vertical external-cavity surface-emitting lasers (VECSEL)~\cite{tropper04,LMB-OE-10} and injected passive high-Q Kerr resonators such as fiber loops~\cite{LCK-NAP-10} and microrings~\cite{HBJ-NAP-14}. The physical processes of the OFC formation in Kerr resonators depend on the respective signs of the dispersion and Kerr nonlinearity and they are well modeled by the Lugiato-Lefever equation or coupled mode models, see, e.g.,~\cite{CSY-PRL-10,CRSE-OL-13,PPR-PR-18,CPP-PRL-21}. In particular, both bright and dark temporal localized states (TLSs) are observed in the anomalous and normal dispersion regimes, respectively~\cite{LCK-NAP-10,LLK_OE_15,XXL_NP_15}. Such TLSs can be multistable and exhibit variable widths. They can be build up from fronts or domain walls that interlock and connect domains of the corresponding lower and higher continuous wave (CW) background intensities~\cite{ROSANOV-BOOK-96,PKG_PRA_16,GWM_EPJD_17}. The coexistence between dark and bright TLSs is permitted by the presence of third order dispersion (TOD)~\cite{PGL-OL-14,PGG_PRA_17}. The locking of domain walls was also demonstrated in doubly resonant dispersive parametric oscillators~\cite{PGH-OL-19} and cavity-enhanced second harmonic generation~\cite{ARHGL-OL-20}.

Recently, an alternative method for the generation of phase-locked OFCs with tunable repetition rate that maintain the high optical power levels characteristic of VECSELs was proposed using a first principle model relying on delay algebraic equations (DAEs)~\cite{SPV-OL-19}. There, dark and bright TLSs form via connecting fronts between bistable CW background states. The latter would interlock at multiple equilibrium distances leading to a rich ensemble of multistable solutions. However, the study performed in ~\cite{SPV-OL-19} was impeded by the inherent difficulties of studying DAEs. This problem was circumvented by studying an alternative singularly perturbed model, which imposed strong limits on the analysis to short cavity lengths and low finesses.
\begin{figure}[t!]
	\centering
	\includegraphics[width=1\columnwidth]{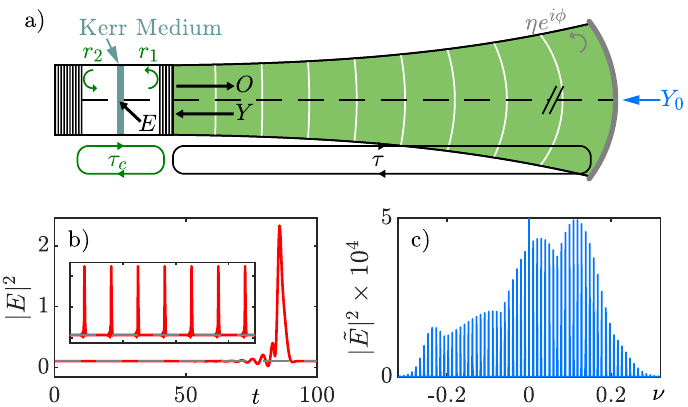}
	\caption{(a) Schematic of micro-cavity containing a Kerr medium coupled to an external cavity which is closed by a mirror with reflectivity $\eta$. It is driven by a monomode injection field with amplitude $Y_0$. (b) Single TLS circulating in the external cavity obtained by integrating Eqs.~(\ref{eq:DAE1}-\ref{eq:DAE2}). The grey dashed line corresponds to the value of the lower CW state. The inset shows the dynamics over several round-trips. (c) Frequency spectrum of the TLS solution shown in (b) where the injection frequency was cut for clarity. The parameters are $(Y_0,\delta,h,\eta,\varphi,\tau)=(0.6,1.5,2,0.75,0,100)$.}
	\label{fig:setup}
\end{figure} 

In this paper we unveil the mechanisms responsible for the domain walls locking in this system. A multiple time-scale analysis performed in the vicinity of the critical point associated with the onset of optical bistability shows that the slow evolution of the system is captured by a real cubic Ginzburg-Landau equation. Our results are compared with a bifurcation analysis that is free from the approximations and the limits imposed by the singular perturbation approach used in ~\cite{SPV-OL-19}.

A schematic setup of the system is presented in Fig.~\ref{fig:setup}~(a). It is composed of a disk-shaped monomode micro-cavity similar to a VCSEL structure but that contains a nonlinear Kerr medium. The cavity is a few micrometer long and its radius may scale up to $100\,\mu$m. The micro-cavity is closed by two distributed Bragg mirrors with reflectivities $r_{1,2}$, whereas the long external cavity with round-trip time $\tau\gg \tau_c$ is closed by a feedback mirror with reflectivity $\eta$ and the feedback phase $\phi$. We consider external cavity lengths of a few centimeters and picosecond pulsewidth thus giving rise to OFC with THz bandwidth and GHz free spectral range.
Trains of short optical pulses with periodicity $\gtrsim\tau$, as shown in the inset of Fig.~\ref{fig:setup}~(b), can be generated for a range of the injection strength $Y_0$ and of the detuning $\delta=\omega_c-\omega_0$, where  $\omega_c$ and $\omega_0$ are the microcavity resonance and injection frequencies, respectively. This temporal dynamics gives rise to an OFC with a width proportional to the inverse of the photon lifetime in the microcavity and a free spectral range of $\tau^{-1}$, see Fig.~\ref{fig:setup}~(c). The external phase $\varphi=\phi+\omega_0\tau$ is the sum of the propagation phase in the external cavity and of the phase shift induced by the feedback mirror. It corresponds to the detuning between the injection and the nearest external cavity mode. 

The evolution of the (normalized) slowly varying field envelopes in the micro-cavity $E$ and the external cavity $Y$ is governed by~\cite{SPV-OL-19}
\begin{align}
\dot{E} &= \left[-1 + i \left(|E|^2-\delta\right)\right] E + h Y \label{eq:DAE1}\,,\\
Y &= \eta e^{i\varphi} \left[E(t-\tau) - Y(t-\tau)\right] + \sqrt{1-\eta^2}Y_0\,. \label{eq:DAE2}
\end{align}
The Eqs.~(\ref{eq:DAE1}-\ref{eq:DAE2}) were obtained from first principles and by solving exactly the field equations in the linear parts of the micro-cavity, connecting the
fields at the interface with the nonlinear medium as a boundary condition. The approach is similar to the one used for modeling VCSEL-RSAM mode-locked lasers, see e.g., \cite{SCM-PRL-19,SHJ-PRAp-20,HGJ-OL-21,SGJ-PRL-22}.
The output $O=E - Y$ is the combination of the intra-cavity photons transmitted and reflected by the micro-cavity and it is re-injected after attenuation and a time delay $\tau$. The coupling between the intra- and external cavity fields is given by a DAE ~(\ref{eq:DAE2}) which takes into account all the multiple reflections in a possibly high finesse external cavity. The light coupling efficiency in the cavity is given by the factor $h=\left(1+|r_2|\right)\left(1-|r_1|\right)/\left(1-|r_1||r_2|\right)$. For a perfectly reflecting bottom mirror $r_2=1$ and, hence, $h=2$, which corresponds to the imbalanced Gires–Tournois interferometer regime~\cite{GT-CRA-64}. Group delay dispersion (GDD) and third order dispersion are naturally captured by Eqs.~(\ref{eq:DAE1}-\ref{eq:DAE2})~\cite{SCM-PRL-19,SPV-OL-19}. The amount and the sign of the GDD (which is typically the dominating effect outside resonance) are tunable by choosing the frequency of operation with respect to the cavity resonance. However, TOD becomes the leading term  around resonance as the second-order contribution vanishes and switches sign. Due to TOD, the resulting temporal pulses can possess strong oscillatory tails, see  Fig.~\ref{fig:setup}~(b), where the typical time trace of a single TLS obtained from integrating numerically Eqs.~(\ref{eq:DAE1}-\ref{eq:DAE2}) is shown. As a result, the envelope of the corresponding OFC is asymmetrical (cf. Fig.~\ref{fig:setup}~(c)). 
\begin{figure}[t!]
	\centering
	\includegraphics[width=1\columnwidth]{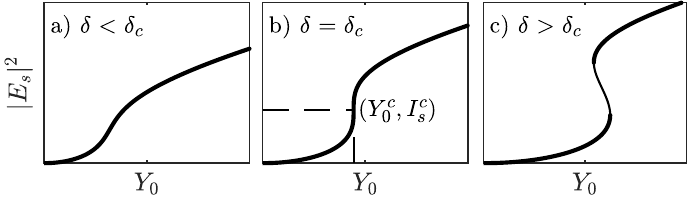}
	\caption{Branch of the CW solution represented by the constant microcavity field intensity $I_s=|E_s|^2$ as a function of the injection amplitude $Y_0$ for the detuning values below (a), at (b) and above (c) the onset of the optical bistability located at $(Y_0^c,\,I_s^c)$. Stable (unstable) solutions are indicated with thick (thin) lines.}
	\label{fig:unfolding}
\end{figure}
The Kerr nonlinearity causes a bistable CW response in a certain range of the injection intensity and of the detuning. 
We consider the situation discussed in \cite{SPV-OL-19} in which the injection was set to be resonant with an external cavity mode, i.e. $\varphi=0$. 
In this situation, the steady state CW solutions $I_s=|E_s|^2$ are obtained solving Eqs.~(\ref{eq:DAE1}-\ref{eq:DAE2}) for $Y_0$ and $\varphi=0$ as
\begin{equation}
Y_{0}^{2}=\frac{k^{2}+\left(I_{s}-\delta\right)^{2}}{h^{2}\frac{1-\eta}{1+\eta}}I_{s}\,, \label{eq:cw}
\end{equation} 
where $k=\dfrac{1+\left(1-h\right)\eta}{1+\eta}$. The critical detuning $\delta_c=\sqrt{3}k$  corresponds to the onset of the optical bistability for some values of $(Y_0^c,\,I_s^c)$. For $\delta<\delta_c$ the system is monostable whereas for $\delta>\delta_c$ it shows a bistable response, see Fig.~\ref{fig:unfolding}.


We used a recently developed extension of the DDE-BIFTOOL~\cite{DDEBT} package that allows for the bifurcation analysis of algebraic and neutral delayed equations~\cite{BKW_JDEA_06,SHJ-PRAp-20,HGJ-OL-21,SGJ-PRL-22}.
%
\begin{figure}[b!]
	\centering
	\includegraphics[width=1\columnwidth]{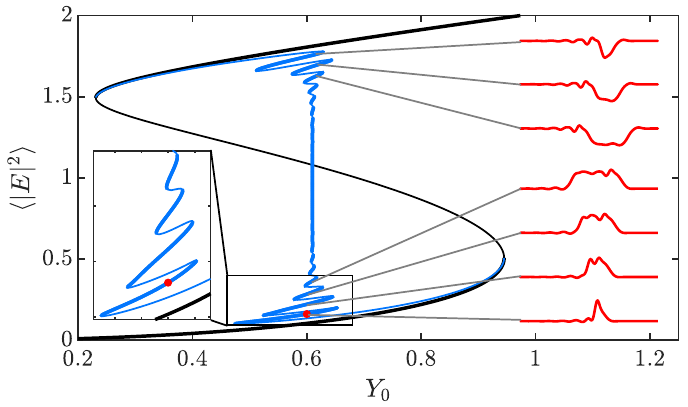}
	\caption{Bifurcation diagram of Eqs.~(\ref{eq:DAE1}-\ref{eq:DAE2}). The CW and the TLS branches are depicted in black and blue while thick (thin) lines denote stable (unstable) solutions, respectively. On the right, solution profiles at the position indicated by the grey lines are shown. The red dot corresponds to the parameters of Fig.~\ref{fig:setup}~(b,c). Left inset: zoom on the lower part of the branch.}
	\label{fig:snaking}
\end{figure}
We present in Fig.~\ref{fig:snaking} the resulting bifurcation diagram for the same parameter set used in Fig.~\ref{fig:setup}~(b,c)  (cf. red point in Fig.~\ref{fig:snaking}). Note that for this parameter set, the detuning is far away from $\delta_c$ which leads to a wide CW bistability range, depicted in black.
\begin{figure*}[t!]
	\centering
	\includegraphics[width=1\textwidth]{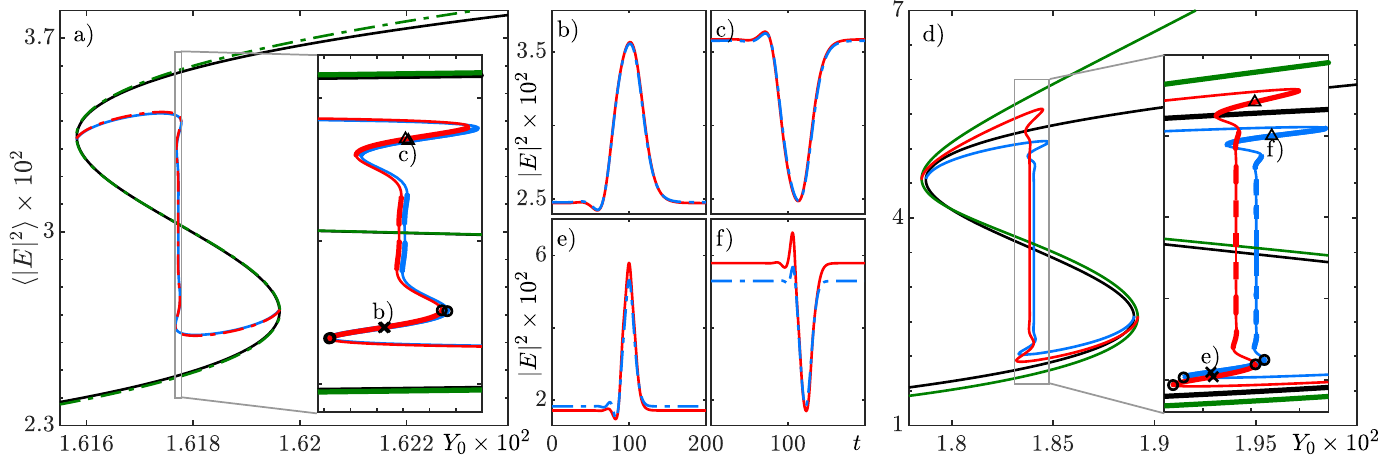}
	\caption{(a,d): Integrated intensity of the $E$-field of the full DAE~(\ref{eq:DAE1}-\ref{eq:DAE2}) as a function of the injection $Y_0$ superposed with the solutions of the normal form PDE~(\ref{eq:PDE}) obtained for $\delta-\delta_c=$ (a) $10^{-3}$, (d) $9\times 10^{-3}$. The CW and TLSs branches of the DAE (PDE) are depicted in black and blue (green and red), respectively. For clarity, stability information is only shown in the insets which give a zoomed view on the snaking region. Here, thick (thin) lines denote stable (unstable) solutions. (b,c,e,f): Solution profiles of the DAE (blue dashed) and the normal form PDE (red solid) at the locations marked in the insets in (a),(d) by crosses (bright TLSs, cf. (b,e)) and triangles (dark TLSs, cf. (c,f)). The black circles denote the folds limiting the stability of the bright TLS (cf. Fig.~\ref{fig:two_param_diag}). The parameters are $(h,\eta,\varphi)=(2,0.95,0)$, $\delta_c=0.044$, and $\tau=200$.}
	\label{fig:compare}
\end{figure*}
From the CW branch, the branch of the TLSs emerges in a subcritical Adronov-Hopf bifurcation; stable TLSs of different widths are connected by bridges of unstable solutions. The insets on the right in Fig.~\ref{fig:snaking} show the solution profiles at different levels along the snaking branch. One can see that each TLS is followed by a decaying oscillating tail induced by TOD. Due to interactions, the drifting fronts can only lock at discrete distances and for a certain injection range. This results in the snaking of the solution branch. Note that at the Maxwell point $Y_{0,MP}\approx 0.61$, the two fronts possess the same speed and their dynamics is arrested.

Althought quantitatively predictive, the bifurcation analysis presented in Fig.~\ref{fig:snaking} does not provide an intuitive interpretation of the dynamics occurring. For instance, the role of the GDD and of the TOD in the DAEs (\ref{eq:DAE1}-\ref{eq:DAE2}) is hidden. In order to gain a better understanding of the underlying physical mechanisms, we transform the DAE model (\ref{eq:DAE1}-\ref{eq:DAE2}) into a partial differential equation (PDE) using a rigorous multiple time scale analysis as in ~\cite{SCM-PRL-19,SHJ-PRAp-20,SGJ-PRL-22}.
We use the onset of bistability $(Y_0^c,\,I_s^c)$ at $\delta=\delta_c$ as an expansion point, i.e. we define deviations from the critical values $\theta=\delta-\delta_c,\, z_0=Y_0-Y_0^c$ and $u+iv=E-E_s^c$. We stress that accurate predictions can be obtained from the PDE only in the vicinity of the onset of optical bistability. As the TLSs exist in the long cavity limit, we define a smallness parameter $\epsilon=1/\tau\ll 1$ in which the deviations are expanded. Further, we define multiple time scales in powers of $\epsilon$ as the dynamics of the TLSs can be separated into a slow time scale $\xi$ governing the evolution between round-trips and a fast time scale $t$ describing the dynamics within one round-trip~\cite{GP-PRL-96}. After inserting all expansions in the original model~(\ref{eq:DAE1}-\ref{eq:DAE2}), one can solve the resulting equations order by order, see Supplementary Material for details. The resulting equation governing the dynamics of the real part of the deviation of the electric field  $u=u(t,\,\xi)$ is
\begin{align}
\partial_\xi u =&  d_2 \partial_t^2 u+ d_3 \partial_t^3 u + \varrho u^2 \partial_t u+f\left(u\right) \,. \label{eq:PDE}
\end{align} 
The diffusion and TOD coefficients $d_j=d_j(\eta,h)$, as well as the coefficient $\varrho$ of the nonlinear drift term read
\begin{align}
\begin{split}
d_2 &= \frac{\omega_1^2}{2}\frac{1-\eta}{1+\eta},\,\qquad d_3=-\frac{|\omega_1|^3}{3}\frac{1-\eta\left(1-\eta\right)}{\left(1+\eta\right)^{2}}\,,\\
\varrho &=\frac{4|\omega_1|}{\sqrt{3}}\frac{\left(1+\eta\right)\left(1-\eta\right)}{h\eta},\,\quad \omega_{1}=-\dfrac{\left(1+\eta\right)^{2}}{\eta h}\,.
\end{split}
\end{align} 
Equation~(\ref{eq:PDE}) is a real Ginzburg-Landau equation with TOD, nonlinear drift and a cubic nonlinearity $f\left(u\right)=c_0+c_1u+c_2u^2+c_3u^3$. Its ingredients are particularly instructive and tell us how each of the physical effects is influenced by the system parameters. In particular, the diffusion and TOD coefficients $d_j$ are functions of the reflectivity $\eta$ and of the light coupling efficiency $h$. The coefficients of the nonlinearity are 
\begin{align}
c_0=&\frac{1+\eta}{\eta}\sqrt{1-\eta^{2}}\left[Y_{0}-Y_{0}^{c}\right] \nonumber\\
 &+\omega_{1}\left[\delta-\delta_{c}\right]\left(\sqrt{\frac{k}{2\sqrt{3}}}+\frac{1}{8}\sqrt{\frac{\sqrt{3}}{2k}}\left[\delta-\delta_{c}\right]\right)\,,\\
c_1 =& \left|\omega_1\right| \frac{\delta-\delta_{c}}{\sqrt{3}},\quad c_2 = \left|\omega_1\right| \frac{\delta-\delta_{c}}{\sqrt{6\sqrt{3}k}},\quad c_3=-\frac{4|\omega_1|}{3\sqrt{3}}\,\nonumber.
\end{align} 
In the good cavity limit $\eta\rightarrow 1$, $d_2$ vanishes and the dynamics is entirely controlled by TOD, which explains the strongly oscillating tails observed for large values of $\eta$. In this regime, the nonlinear drift becomes negligible as $\varrho$ vanishes. In contrast, the coefficients $c_i$ are controlled by detuning, injection, and the distance from the onset of bistability. Hereby they define the values of the CW solutions and the width of the bistability region. The tails of appearing fronts connecting these CW states are however controlled by the spatial operator and ultimately the values of $(\eta,h)$. This, in turn, affects the front interaction and the possible amount of TLSs that may appear within the snaking region. Note that in~\cite{HBC-PRA-19}, an amplitude equation for a photonic crystal fiber resonator was derived close to the onset of bistability. There, starting form a generalized Lugiato-Lefever equation in the anomalous dispersion regime, a Swift-Hohenberg model with TOD was obtained. In our case, however, as we operate with normal dispersion ($\delta>0$), $d_2>0$  making further expansion of the spatial operator to fourth order unnecessary.

We compare the solutions of the normal form ~\eqref{eq:PDE} with those obtained from the full DAE model (\ref{eq:DAE1}-\ref{eq:DAE2}); our results are presented in Fig.~\ref{fig:compare}. Since \eqref{eq:PDE} only governs the real part of $E$, we reconstructed the full electric field as $E=E_s^c+u+iv$, where $v$ is obtained from the multiscale analysis and $E_{s}^c$ is obtained from \eqref{eq:cw}. Figure~\ref{fig:compare}~(a),(d) show branches of the CW solution and TLSs, obtained from the full DAE model (black, blue) and the PDE normal form (green, red), respectively, for different deviations from the critical detuning $\delta_c$. For the small deviation (cf. Fig \ref{fig:compare}~(a)), one observes an excellent agreement between the branches (see the inset). In Fig.~\ref{fig:compare} (b), [(c)], we compare the profiles of a bright [dark] TLS marked on the branch by a cross [triangle], respectively. Again, the agreement is excellent. Moving further away from the onset (cf. Fig.~\ref{fig:compare} (d)), the agreement between the DAE and PDE models is still qualitatively good but the branches deviate from each other at high intensities or when looking at the profiles in Fig.~\ref{fig:compare} (e) and (f), respectively.

\begin{figure}
	\centering
	\includegraphics[width=1\columnwidth]{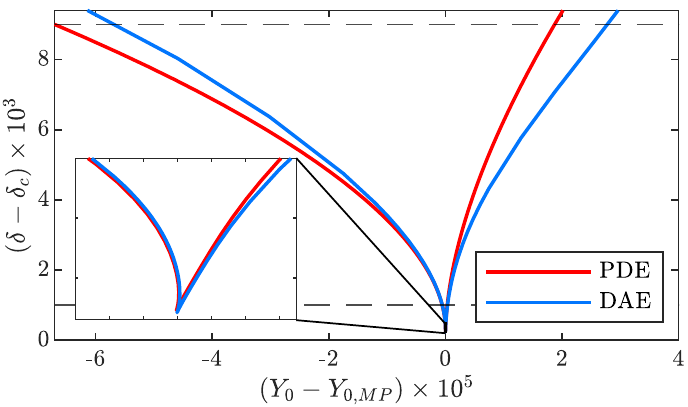}
	\caption{Evolution of the folds limiting the stability of the bright TLS (cf. circles in Fig. \ref{fig:compare} (a),(d)) for the full system described by the DAE in Eqs. (\ref{eq:DAE1}-\ref{eq:DAE2}) (blue) and the developed normal form PDE (\ref{eq:PDE}) (red) in the two-parameter plane $(Y_0,\delta)$. The horizontal dashed lines correspond to the detuning values for which Fig. \ref{fig:compare} (a) and (d) are shown. For clarity, the injection values are shifted by $Y_{0,MP}$ which is a function of the other parameters (in particular, of $\delta$) giving an approximate analytical solution for the Maxwell point. Parameters as in Fig. \ref{fig:compare}.}
	\label{fig:two_param_diag}
\end{figure}

The agreement between the PDE (\ref{eq:PDE}) and the DAE (\ref{eq:DAE1}-\ref{eq:DAE2}) close to the bistability onset can be best observed when looking at the two-parameter-plane $(Y_0,\,\delta)$ in Fig.~\ref{fig:two_param_diag}, where the evolution of the folds limiting the stability of the bright TLS are shown (cf. the black circles in the insets of Fig.~\ref{fig:compare}). Note that here the injection values are shifted by the value $Y_{0,MP}=\sqrt{\frac{1}{27}\frac{(\delta+\delta_c)^{3}}{h^{2}\frac{1-\eta}{1+\eta}}}$ for clarity. This expression gives an approximate value of the Maxwell point by evaluating the mean value between the folds of the CW branch. Thus, the correction accounts for the shift of the Maxwell point for increasing $\delta$ such that $Y_0-Y_{0,MP}$ is centered around zero. Naturally, the agreement is best close to the onset of bistability where the two folds merge (see inset).

In summary, we discussed the mechanisms responsible for the domain walls locking and the formation of temporal localized structures and optical frequency combs in vertical external-cavity Kerr-Gires–Tournois interferometers. Using a first principle model relying on delay algebraic equations and by means of a rigorous multiple-scale analysis performed at the onset of optical bistability, we unveiled a minimal scenario of pattern formation that is captured by a real Ginzburg-Landau equation with higher-order effects. 
We believe~\eqref{eq:PDE} to be of wider applicable to other systems close to the onset of bistability and where high-order effects are responsible for domain wall dynamics.

\section*{Acknowledgment}
T.G.S. thanks the foundation “Studienstiftung des deutschen Volkes” for financial support, J.J. acknowledges funding of the MINECO Project MOVELIGHT (PGC2018-099637-B-100 AEI/FEDER UE).

\section*{Disclosures}
The authors declare no conflicts of interest

See Supplement 1 for supporting content


\end{document}


\maketitle

\section{The Onset of Bistability}
The DAE model  for the formation of temporal localized states (TLSs) and optical frequency combs (OFCs) in a micro-cavity operating in the Gires–Tournois regime containing a nonlinear Kerr medium and coupled to an long external cavity under CW injection reads
\begin{align}\dot{E} & =\left[-1+i\left(|E|^{2}-\delta\right)\right]E+hY\,, \label{eq:E}\\
Y & =\eta e^{i\varphi}\left[E(t-\tau)-Y(t-\tau)\right]+\sqrt{1-\eta^{2}}Y_{0}\,. \label{eq:Y}
\end{align}
The steady state continuous wave (CW) solutions $I_s=|E_s|^2$ of Eqs.~(\ref{eq:E}-\ref{eq:Y})  can be easily obtained for the case $\varphi=0$ and read
\[
Y_{0}^{2}=\frac{k^{2}+\left(I_{s}-\delta\right)^{2}}{h^{2}\frac{1-\eta}{1+\eta}}I_{s},
\]
where $k=\dfrac{1+\left(1-h\right)\eta}{1+\eta}$. For certain system parameters, the curve $I_s(Y_0)$ exhibits bistabile response. In fact, the onset of bistability can
be calculated analytically. It is defined as the value of the critical detuning $\delta_{c}$
for which the steady state solutions $I_{s}$ possesses a saddle point,
i.e., $\dfrac{\partial Y_{0}}{\partial I_{s}}=\dfrac{\partial^{2}Y_{0}}{\partial I_{s}^{2}}=0$.
At this point, we find critical values for the detuning $\delta_{c}$, the injection
$Y_{0}^{c}$ and the electric field $E_{s}^{c}$:
\begin{align}
	\delta_{c} & =\sqrt{3}k, \quad Y_{0}^{c} =\frac{2k}{3}\sqrt{\frac{2\sqrt{3}}{h^{2}\frac{1-\eta}{1+\eta}}k},\quad E_{s}^{c}=\sqrt{\frac{3k}{2\sqrt{3}}}\left(1-i\frac{1}{\sqrt{3}}\right)\,.
\end{align}

\section{Expansions}
Next, we apply a multiple time scale method to the periodic solution of Eqs.~(\ref{eq:E}-\ref{eq:Y}). For this purpose, we normalize the long delay time $\tau$ and define an smallness parameter $\epsilon=\tau^{-1}$ which yields
\begin{eqnarray}
	\varepsilon\dot{E} & = & \left[-1+i\left(\left|E\right|^{2}-\delta\right)\right]E+hY\,,\\
	Y & = & \eta\left[E\left(t-1\right)-Y\left(t-1\right)\right]+\sqrt{1-\left|\eta\right|^{2}}Y_{0}\,.
\end{eqnarray}
We define new variables $e$ and $y$ as small deviations from the steady
states at the critical point and separate into real and imaginary parts: $E=E_{s}^{c}+e=E_{sx}^{c}+iE_{sy}^{c}+e_{x}+ie_{y}$
and $Y=Y_{s}^{c}+y=Y_{s}^{c}+y_{x}+iy_{y}$. Further, we define $\theta$ and
$z_{0}$ are the deviation from the critical parameter values for
detuning and injection, respectively: $\delta=\delta_{c}+\theta$
and $Y_{0}=Y_{0}^{c}+z_{0}$. Then the equations for the fields deviations read
\begin{align}
	\begin{split}
		\epsilon\frac{de_{x}}{dt}  =&-\sqrt{\frac{k}{2\sqrt{3}}}\theta-\left(1-k\right)e_{x}+\theta e_{y}+hy_{x}-\sqrt{2\sqrt{3}k}e_{x}e_{y}\\
	&+\sqrt{\frac{k}{2\sqrt{3}}}e_{x}^{2}+\sqrt{\frac{3\sqrt{3}k}{2}}e_{y}^{2}-e_{x}^{2}e_{y}-e_{y}^{3}\,,
	\end{split}
	\label{eq:dex_dt}\\
	\begin{split}
	\epsilon\frac{de_{y}}{dt}  =&-\sqrt{\frac{\sqrt{3}k}{2}}\theta+\left(\frac{2}{\sqrt{3}}k-\theta\right)e_{x}-\left(1+k\right)e_{y}+hy_{y}-\sqrt{\frac{2k}{\sqrt{3}}}e_{x}e_{y}\\
	&+3\sqrt{\frac{\sqrt{3}k}{2}}e_{x}^{2}+\sqrt{\frac{\sqrt{3}k}{2}}e_{y}^{2}+e_{x}e_{y}^{2}+e_{x}^{3}\,,
	\end{split}
	\label{eq:dey_dt}\\
	y_{x} =&\eta\left(e_{x}\left(t-1\right)-y_{x}\left(t-1\right)\right)+\sqrt{1-\left|\eta\right|^{2}}z_{0}\,,\label{eq:yx}\\
	y_{y} =&\eta\left(e_{y}\left(t-1\right)-y_{y}\left(t-1\right)\right)\,.\label{eq:yy}
\end{align}
We define new time scales $t_{0}=\omega t$, $t_{1}=0$, $t_{2}=\epsilon^{2}t$
and $t_{3}=\epsilon^{3}t$, where small changes in $\omega$ account
for the drift: $\omega=1+\epsilon\omega_{1}+\epsilon^{2}\omega_{2}+\epsilon^{3}\omega_{3}+\dots$.
Therefore, the time derivative is $\dfrac{d}{dt}=\mathcal{T}_{0}+\epsilon\mathcal{T}_{1}+\epsilon^{2}\mathcal{T}_{2}+\epsilon^{3}\mathcal{T}_{3}+\dots$,
where $\mathcal{T}_{0}=\partial_{\text{0}}$, $\mathcal{T}_{1}=\omega_{1}\partial_{0}$,
$\mathcal{T}_{2}=\omega_{2}\partial_{0}+\partial_{2}$ and $\mathcal{T}_{3}=\omega_{3}\partial_{0}+\partial_{3}$.
Further, the delayed fields are now functions of the new time scales,
i.e., $\Lambda\left(t-1\right)=\Lambda\left(t_{0}-\omega,t_{2}-\epsilon^{2},t_{3}-\epsilon^{3}\right)$
for $\Lambda=e_{x},e_{y},y_{x},y_{y}$. This expression needs to be
expanded in powers of $\epsilon$:
\begin{align}
	\Lambda\left(t-1\right) & =\left(\mathcal{S}_{0}+\epsilon\mathcal{S}_{1}+\epsilon^{2}\mathcal{S}_{2}+\epsilon^{3}\mathcal{S}_{3}+\dots\right)\Lambda\left(t_{0}-1,t_{2},t_{3}\right)\,,
\end{align}
where $\mathcal{S}_{0}=1$, $\mathcal{S}_{1}=-\omega_{1}\partial_{0}$,
$\mathcal{S}_{2}=-\omega_{2}\partial_{0}+\frac{\omega_{1}^{2}}{2}\partial_{0}^{2}-\partial_{2}$
and $\mathcal{S}_{3}=-\omega_{3}\partial_{0}-\frac{\omega_{1}^{3}}{6}\partial_{0}^{3}+\omega_{1}\omega_{2}\partial_{0}^{2}+\omega_{1}\partial_{0}\partial_{2}-\partial_{3}$.
Further, we need to define the expantions of the fields and parameters.
First, all fields are expanded as 
\begin{equation}
	\Lambda\left(t\right)=\varepsilon\Lambda^{\left(1\right)}\left(t_{0},t_{2},t_{3}\right)+\varepsilon^{2}\Lambda^{\left(2\right)}\left(t_{0},t_{2},t_{3}\right)+\cdots\,.
\end{equation}
Finally, we also expand the variations to the critical parameter values
\begin{align}
	\theta & =\epsilon\theta_{1}+\epsilon^{2}\theta_{2}+\epsilon^{3}\theta_{3}+\epsilon^{4}\theta_{4}+\dots\,,\\
	z_{0} & =\epsilon z_{0}^{\left(1\right)}+\epsilon^{2}z_{0}^{\left(2\right)}+\epsilon^{3}z_{0}^{\left(3\right)}+\epsilon^{4}z_{0}^{\left(4\right)}+\dots\,.
\end{align}
We insert all expansions and scalings into Eqs. (\ref{eq:dex_dt})-(\ref{eq:yy}) and sort for orders
in $\epsilon$. As we stretched the $t_{0}$-scale in such a way that
the drift is cancelled, we can assume that $e$ and $y$ are 1-periodic, i.e.,
$\Lambda\left(t_{0}-1\right)=\Lambda\left(t_{0}\right)$ for $\Lambda=e_{x},e_{y},y_{x},y_{y}$.

\section{Solving the Orders}

\subsection*{First Order}
In the first order $\mathcal{O}\left(\epsilon\right)$, the following
linear system yields:
\begin{align}
0=P v^{\left(1\right)}+b\,,\label{eq:orderEq1}
\end{align}
\begin{align}
\text{where } &P=\left(\begin{array}{cccc}
					k-1 & 0 & h & 0\\
					\frac{2k}{\sqrt{3}} & -\left(k+1\right) & 0 & h\\
					\eta & 0 & -\left(1+\eta\right) & 0\\
					0 & \eta & 0 & -\left(1+\eta\right)
					\end{array}\right)\,,\\
&v^{\left(1\right)} = \left(\begin{array}{c}
						e_{x}^{\left(1\right)}\\
						e_{y}^{\left(1\right)}\\
						y_{x}^{\left(1\right)}\\
						y_{y}^{\left(1\right)}
						\end{array}\right), \quad
b=z_{0}^{\left(1\right)}\sqrt{1-\eta^{2}}\left(\begin{array}{c}
0\\
0\\
1\\
0
\end{array}\right)-\sqrt{\frac{k}{2\sqrt{3}}}\theta_{1}\left(\begin{array}{c}
1\\
\sqrt{3}\\
0\\
0
\end{array}\right)	\,.
\end{align}
The homogeneous system $0=Pv^{\left(1\right)}$ has a kernel, i.e.,
the kernel eigenvector $\mu$ is the solution of the linear system:
\begin{equation}
	\mu=\left(\begin{array}{c}
		1\\
		\frac{1}{\sqrt{3}}\\
		\frac{\eta}{1+\eta}\\
		\frac{\eta}{\sqrt{3}\left(1+\eta\right)}
	\end{array}\right).
\end{equation}
The adjoint linear operator $P^{\dagger}$ also has a kernel which
is spanned by the eigenvector 
\begin{equation}
	\nu=\left(\begin{array}{c}
		\frac{1+\eta}{h}\\
		0\\
		1\\
		0
	\end{array}\right).
\end{equation}
This can be used to apply the Fredholm alternative ($Pv^{\left(1\right)}+b=0\Rightarrow\nu^{\dagger}b=0$)
to \ref{eq:orderEq1} in order to find the corresponding solvability
condition which reads
\begin{equation}
	z_{0}^{\left(1\right)}\sqrt{1-\eta^{2}}-\frac{1+\eta}{h}\sqrt{\frac{k}{2\sqrt{3}}}\theta_{1}=0.
\end{equation}
Taking the solution of the inhomogeneous system and the solvability
condition into consideration the full solution of the first order
reads
\begin{equation}
	v^{\left(1\right)}=v_{1}^{\left(1\right)}\left(\begin{array}{c}
		1\\
		\frac{1}{\sqrt{3}}\\
		\frac{\eta}{1+\eta}\\
		\frac{\eta}{\sqrt{3}\left(1+\eta\right)}
	\end{array}\right)+\sqrt{\frac{\sqrt{3}}{2k}}\theta_{1}\left(\begin{array}{c}
		0\\
		-\frac{1}{2}\\
		\frac{k}{h\sqrt{3}}\\
		-\frac{1}{2}\frac{\eta}{1+\eta}
	\end{array}\right).
\end{equation}

\subsection*{Second Order}
In the next second order $\mathcal{O}\left(\epsilon^{2}\right)$ one obtains
\begin{align}
	0 & =Pv^{\left(2\right)}+z_{0}^{\left(2\right)}\sqrt{1-\eta^{2}}\left(\begin{array}{c}
		0\\
		0\\
		1\\
		0
	\end{array}\right)-\mathcal{T}_{0}\left(\begin{array}{c}
		v_{1}^{\left(1\right)}\\
		v_{2}^{\left(1\right)}\\
		0\\
		0
	\end{array}\right)+\eta\mathcal{S}_{1}\left(\begin{array}{c}
		0\\
		0\\
		v_{1}^{\left(1\right)}-v_{3}^{\left(1\right)}\\
		v_{2}^{\left(1\right)}-v_{4}^{\left(1\right)}
	\end{array}\right)+\Theta_{2}+N_{2}\,,\label{eq:OrderEq2}
\end{align}
where $\Theta_{2}$ and $N_{2}$ are the terms stemming from detuning
$\theta e_{x,y}$ and the nonlinearities. The homogeneous system is
the same as before such that we again apply the Fredholm alternative
to obtain the following solvability condition.

\begin{align}
0&=z_{0}^{\left(2\right)}\sqrt{1-\eta^{2}}-\sqrt{\frac{k}{2\sqrt{3}}}\frac{1+\eta}{h}\theta_{2}+\frac{1+\eta}{\sqrt{3}h}\theta_{1}v_{1}^{\left(1\right)}-\frac{1+\eta}{8h}\sqrt{\frac{\sqrt{3}}{2k}}\theta_{1}^{2} \label{eq:second order}
\end{align}
Here, the drift was cancelled by choosing $\omega_{1}=-\frac{\left(1+\eta\right)^{2}}{\eta h}$.
This can be done as no assumptions concerning the scaling factor
$\omega$ of the $t_{0}$-timescale were made such that we can choose the $\omega_{i}$
in a way which is convenient (i.e. to cancel the drift).

Eq. (\ref{eq:second order}) implies that $v_{1}^{\left(1\right)}$ is a constant which depends only on the deviations to critical point and which is not time dependent. This is problematic as we later want to derive a differential equation. However, we remember that we can choose the scaling of the parameter deviations and at this point we find that we have to set $\theta_1=0$ (i.e. $\theta=\mathcal{O}\left(\epsilon^2\right)$) in order to proceed. Then the solvability condition reads
\begin{equation}
	z_{0}^{\left(2\right)}=\sqrt{\frac{1}{2\sqrt{3}}\frac{\eta\left(1-h\right)+1}{1-\eta}}\frac{\theta_{2}}{h}.
\end{equation}
In order to solve the following
orders, one has to compute the full solution including the inhomogeneous
system for the second and third components: 
\begin{align}
	v_{2}^{\left(2\right)} & =\frac{1}{\sqrt{3}}v_{1}^{\left(2\right)}-\frac{1}{2\sqrt{k}}\left[\sqrt{\frac{\sqrt{3}}{2}}\theta_{2}-4\sqrt{\frac{2}{3\sqrt{3}}}\left(v_{1}^{\left(1\right)}\right)^{2}\right],\\
	v_{3}^{\left(2\right)} & =\frac{\eta}{1+\eta}v_{1}^{\left(2\right)}+\frac{1}{1+\eta}z_{0}^{\left(2\right)}\sqrt{1-\eta^{2}}+\frac{1}{h}\partial_{0}v_{1}^{\left(1\right)}.
\end{align}

\subsection*{Third Order}
Following the same procedure, we find the following solvability condition
at third order $\mathcal{O}\left(\epsilon^{3}\right)$: 
\begin{align}
\begin{split}
	\partial_{2}v_{1}^{\left(1\right)}=&\frac{1+\eta}{\eta}\sqrt{1-\eta^{2}}z_{0}^{\left(3\right)}-\frac{1}{2}\frac{\left(1-\eta\right)\left(1+\eta\right)}{\eta h}\omega_{1}\partial_{0}^{2}v_{1}^{\left(1\right)}+\frac{4}{3\sqrt{3}}\omega_{1}\left(v_{1}^{\left(1\right)}\right)^{3}\\
	&+\omega_{1}\sqrt{\frac{k}{2\sqrt{3}}}\theta_{3}-\frac{1}{\sqrt{3}}\omega_{1}\theta_{2}v_{1}^{\left(1\right)}\,,
\end{split}
\end{align}
where the drift was canceled by choosing $\omega_{2}=\frac{\left(1+\eta\right)^{4}}{\eta^{2}h^{2}}=\omega_{1}^{2}$.
We only need to compute the full solution for the third component:
\begin{equation}
	v_{3}^{\left(3\right)}=\frac{\eta}{1+\eta}v_{1}^{\left(3\right)}-\frac{1}{h}\left[-\sqrt{\frac{k}{2\sqrt{3}}}\theta_{3}+\frac{1}{\sqrt{3}}\theta_{2}v_{1}^{\left(1\right)}-\partial_{0}v_{1}^{\left(2\right)}-\omega_{1}\partial_{0}v_{1}^{\left(1\right)}-\frac{4}{3\sqrt{3}}\left(v_{1}^{\left(1\right)}\right)^{3}\right].
\end{equation}

\subsection*{Fourth Order}
Finally, in the forth order $\mathcal{O}\left(\epsilon^{4}\right)$
the solvability condition reads
\begin{align}
\begin{split}
\partial_{3}v_{1}^{\left(1\right)}+\partial_{2}v_{1}^{\left(2\right)}=&-\frac{3\left(1-\eta\right)\left(1+\eta\right)}{2\eta h}\omega_{1}^{2}\partial_{0}^{2}v_{1}^{\left(1\right)}-\frac{1}{2}\frac{\left(1+\eta\right)\left(1-\eta\right)}{\eta h}\omega_{1}\partial_{0}^{2}v_{1}^{\left(2\right)}\\
&+\frac{1}{3}\frac{1+\eta\left(\eta-1\right)}{\left(1+\eta\right)^{2}}\omega_{1}^{3}\partial_{0}^{3}v_{1}^{\left(1\right)}-\frac{4}{3\sqrt{3}}\frac{\left(1+\eta\right)\left(1-\eta\right)}{h\eta}\omega_{1}\partial_{0}\left(v_{1}^{\left(1\right)}\right)^{3}\\
&-\omega_{1}\frac{1}{\sqrt{3}}v_{1}^{\left(1\right)}\theta_{3}-\frac{1}{\sqrt{3}}\omega_{1}^{2}\theta_{2}v_{1}^{\left(1\right)}-\omega_{1}\frac{1}{\sqrt{3}}\theta_{2}v_{1}^{\left(2\right)}-\frac{1}{\sqrt{6\sqrt{3}}\sqrt{k}}\theta_{2}\left(v_{1}^{\left(1\right)}\right)^{2}\\
&+\frac{4}{3\sqrt{3}}\omega_{1}^{2}\left(v_{1}^{\left(1\right)}\right)^{3}+\omega_{1}\frac{4}{\sqrt{3}}\left(v_{1}^{\left(1\right)}\right)^{2}v_{1}^{\left(2\right)}+\omega_{1}\frac{1+\eta}{\eta}\sqrt{1-\eta^{2}}z_{0}^{\left(3\right)}\\
&+\frac{1+\eta}{\eta}\sqrt{1-\eta^{2}}z_{0}^{\left(4\right)}+\sqrt{\frac{k}{2\sqrt{3}}}\omega_{1}^{2}\theta_{3}+\frac{1}{8\sqrt{k}}\omega_{1}\sqrt{\frac{\sqrt{3}}{2}}\theta_{2}^{2}+\omega_{1}\sqrt{\frac{\sqrt{3}k}{6}}\theta_{4}\,,
\end{split}
\end{align}
where the drift was canceled by setting $\omega_{3}=\omega_{1}^{3}+\frac{1}{\sqrt{3}}\frac{\left(1+\eta\right)\left(1-\eta\right)}{h\eta}\omega_{1}\theta_{2}$.

\section{Building the Normal Form}
Next, we define an effective slow time scale as $\partial_{\xi}=\varepsilon^{2}\partial_{2}+\varepsilon^{3}\partial_{3}$
and evaluate
\begin{align}
	\partial_{\xi}e_{x} & =\left(\varepsilon^{2}\partial_{2}+\varepsilon^{3}\partial_{3}\right)\left(\epsilon v_{1}^{\left(1\right)}+\epsilon^{2}v_{1}^{\left(2\right)}\right)+\mathcal{O}\left(\epsilon^{5}\right)\\
	& =\epsilon^{3}\partial_{2}v_{1}^{\left(1\right)}+\epsilon^{4}\left(\partial_{3}v_{1}^{\left(1\right)}+\partial_{2}v_{1}^{\left(2\right)}\right)+\mathcal{O}\left(\epsilon^{5}\right)\,,
\end{align}
where we can insert the solvabiliy conditions found before. Next,
we rescale the space with $\epsilon$ such that $\partial_{0}=\frac{1}{\epsilon}\partial_{t}$.
Then, we are in a good position to find all necessary terms to rebuild $e_{x}=\epsilon v_{1}^{\left(1\right)}+\epsilon^{2}v_{1}^{\left(2\right)}+...$
and the parameters $\theta=\delta-\delta_{c}$ and $z_{0}=Y_{0}-Y_{0}^{c}$
up to $\mathcal{O}\left(\epsilon^{5}\right)$. Finally, we obtain the following amplitude equation:
\begin{align}
\begin{split}
\partial_{\xi}e_{x}=&\frac{\omega_{1}^{2}}{2}\frac{1-\eta}{1+\eta}\partial_{t}^{2}e_{x}-\frac{\left|\omega_{1}\right|^{3}}{3}\frac{1-\eta\left(1-\eta\right)}{\left(1+\eta\right)^{2}}\partial_{t}^{3}e_{x}+\frac{4\left|\omega_{1}\right|}{3\sqrt{3}}\frac{\left(1+\eta\right)\left(1-\eta\right)}{h\eta}\partial_{t}e_{x}^{3}\\
&+\frac{1+\eta}{\eta}\sqrt{1-\eta^{2}}\left[Y_{0}-Y_{0}^{c}\right]+\omega_{1}\left[\delta-\delta_{c}\right]\left(\sqrt{\frac{k}{2\sqrt{3}}}+\frac{1}{8}\sqrt{\frac{\sqrt{3}}{2k}}\left[\delta-\delta_{c}\right]\right)\\
&-\frac{\delta-\delta_{c}}{\sqrt{3}}\omega_{1} e_{x}-\omega_{1}\frac{\delta-\delta_{c}}{\sqrt{6\sqrt{3}k}}e_{x}^{2}-\frac{4\left|\omega_{1}\right|}{3\sqrt{3}}e_{x}^{3}+\mathcal{O}\left(\epsilon\right).
\end{split}
\end{align}

\section{Rebuilding the full Field}
The full  complex $E$ field can be reconstructed via $E=E_{sx}^{c}+iE_{sy}^{c}+e_{x}+ie_{y}$ ($=E_s^c+u+iv$ in the notation of the paper)
where $e_{y}$ is given by $e_{y}=\epsilon v_{2}^{\left(1\right)}+\epsilon^{2}v_{2}^{\left(2\right)}+\epsilon^{3}v_{2}^{\left(3\right)}+\mathcal{O}\left(\epsilon^{4}\right)$.
Inserting the solutions from each order yields
\begin{align}
\begin{split}
e_{y}=&\frac{1}{\sqrt{3}}e_{x}-\frac{2}{3k}\left(\delta-\delta_{c}\right)e_{x}+\frac{2}{3}\sqrt{\frac{2\sqrt{3}}{k}}e_{x}^{2}+\frac{8}{9k}e_{x}^{3}-\frac{1}{6}\sqrt{\frac{2\sqrt{3}}{k}}\left(\delta-\delta_{c}\right)\\
&-\frac{\sqrt{3}h}{6k}\sqrt{\frac{1-\eta}{1+\eta}}\left(Y_{\text{0}}-Y_{0}^{c}\right)+\mathcal{O}\left(\epsilon^{4}\right).
\end{split}
\end{align}